\documentclass[a4paper]{jpconf}
\usepackage{graphicx,epsf}
\begin{document}
\title{Toward Radiation-Magnetohydrodynamic Simulations in Core-Collapse Supernovae}
\author{Kei Kotake, Naofumi Ohnishi\dag, Shoichi Yamada\ddag, and Katsuhiko 
Sato$\star$}
\address{ \ Science \& Engineering, Waseda University, 3-4-1 Okubo,
Shinjuku, Tokyo, 169-8555, Japan}
\address{\dag Department of Aerospace Engineering, Tohoku University,
6-6-01 Aramaki-Aza-Aoba, Aoba-ku, Sendai, 980-8579, Japan}
\address{\ddag Advanced Research Institute for Science and Engineering, Waseda University, 3-4-1 Okubo, Shinjuku,
Tokyo, 169-8555, Japan}
\address{$\star$ Department of Physics, School of Science, the University of
Tokyo, 7-3-1 Hongo, Bunkyo-ku, Tokyo 113-0033, Japan}
\begin{abstract}
We report a current status of our radiation-magnetohydrodynamic code for the study of core-collapse supernovae. 
 In this contribution, we discuss the accuracy of our newly developed numerical code 
by presenting the test problem in a static background model.  We also present 
the application to the spherically symmetric core-collapse 
simulations.   
Since close comparison 
with the previously published models is made, we are now applying it for 
the study of magnetorotational core-collapse supernovae. 
\end{abstract}
\section{Introduction}
Recently, multidimensional studies and simulations of core-collapse 
supernovae have come into blossom again since 1990's,
when the direct observations of global asymmetry in SN 1987A were reported.
This trend may be partly because spherically symmetric
supernova simulations have not yet produced explosions
(e.g., \cite{1D})
 albeit with the 
sophistication of the neutrino-transport method including the 
state-of-the-art nuclear physics,  and 
the detailed weak interaction rates. 
Meanwhile, there have been 
the accumulating observations after SN 1987A implying that the
core-collapse supernovae are generally aspherical
(e.g., \cite{wang02}). The degree of asymmetry also 
rises as a function of time, 
which has been interpreted as an evidence that the inner portion of the
explosion are strongly aspherical.

A leap beyond the spherical models seems indeed meaningful, because 
asphericities in the supernova core should have influence 
on the nucleosynthesis, 
neutrino and gravitational-wave emissions \cite{kotake_rev}.
Before the advent of the observations using neutrinos and gravitational waves as new eyes, one hopes to clarify the explosion mechanism of 
aspherical supernovae.

So far, many physical ingredients to produce aspherical explosion have been 
investigated, such as
 convection, possible density inhomogeneities formed prior to core-collapse,
rotation and magnetic fields \cite{ober} (see \cite{kotake_rev} for a collective reference),
 more recently standing shock instability \cite{SASI} and the excitations of g-modes in the protoneutron star \cite{burrows}.
Whatsoever the origin of the asphericity, the neutrino heating 
mechanism should play a key role to drive explosions.
 Thus multidimensional neutrino transport simulations are indispensable.
In this contribution, we report the current status of our 
radiation transport calculations.
\section{Basic equations of radiation-magnetohydrodynamics}
Magnetohydrodynamic (MHD) equations including the neutrino-matter coupling read,
\begin{equation}
\rho \frac{d \mbox{\boldmath$v$}}{dt} + \nabla p + \rho \nabla \Phi - \frac{1}{4 \pi}
(\nabla \times \mbox{\boldmath$B$})\times \mbox{\boldmath$B$} = S_{ M}, 
\label{euler}
\end{equation}
\begin{equation}
\rho \frac{d \displaystyle{\Bigl(\frac{e}{\rho}\Bigr)}}{dt} + p \nabla \cdot \mbox{\boldmath$v$} = S_{ E},
\label{energy}
\end{equation}
\begin{equation}
\frac{dY_e}{dt} = S_{Y_e}. 
\label{ye_flow}
\end{equation}
The left-hand side of the above equations is MHD evolution equations, which we numerically integrate by the Newtonian explicit code ZEUS2D \cite{stone}. Equation (\ref{euler}) is the Euler equation, where $\rho$ is the mass density, $d/dt$ is the Lagrange derivative, $\mbox{\boldmath$v$}$ is the fluid velocity, $p$ is the matter pressure, 
$\mbox{\boldmath$B$}$ is the magnetic fields solved by the
induction equation : $\partial  \mbox{\boldmath$B$}/\partial t = 
\nabla \times( \mbox{\boldmath$v$}\times \mbox{\boldmath$B$})$ keeping the 
field divergence-free ($\nabla \cdot \mbox{\boldmath$B$} = 0$) using the CT scheme, $\Phi$ is the gravitational potential including the general relativistic correction. Equation (\ref{energy}) and (\ref{ye_flow})
 expresses the evolution of specific internal energy ($e$)
 and electron fraction ($Y_e$), respectively. As for the equation of state, 
we incorporate a tabulated one based on relativistic mean field theory 
\cite{shen98}.
   
The right-hand side of the above equations expresses the transfer of  
momentum ($S_{M}$), exchange of energy ($S_{E}$) from neutrinos to matter, and the change of electric charge ($S_{Y_e}$). As for the neutrino reactions, we include the 
so-called standard set denoted in table 1 in \cite{bruenn85}, plus 
nucleon bremsstrahlung \cite{hannestad}. Corrections to the standard neutrino opacities, such as 
the detailed reaction kinematics of nucleon thermal motions,
recoil, and weak magnetism,  pair-annihilation/creations among 
different neutrino flavors, the quenching of the coupling constant above nuclear density, and the modified electron capture rates to the nuclei in the excitation states and so on (see \cite{buras}), we omit them for simplicity in the simulation presented here.  
 To determine the right-hand side of the above equations, one has to know 
the neutrino distribution function $f(t,\mbox{\boldmath$r$}, \mbox{\boldmath$p$})$ by solving the Boltzmann equation with the collisional terms including the above neutrino reactions in its right-hand side. Fully angle-dependent solutions in multidimensional models (even in 2D) are computationally prohibitive,  
we employ the flux limited diffusion approximation to relate the first and zeroth angular moment of the neutrino distribution function, 
\begin{equation}
\mbox{\boldmath$\psi_1$} = \left(\begin{array}{c}
   \psi_{1,r}\\
  \psi_{1,\theta}\\
\end{array}\right)= -  \mbox{\boldmath$\Lambda$}\cdot
\Bigl[\nabla \psi_0 - \mbox{\boldmath$A_1$}\psi_0 - \mbox{\boldmath$C_1$}\Bigr],\label{psi1}
\end{equation}
where $\psi_0$ and $\mbox{\boldmath{$\psi_1$}}$ is the zeroth and first angular moment of the neutrino distribution function, respectively. See equation (A25)  in \cite{bruenn85} for the definition of 
 \mbox{\boldmath$A_1$} and \mbox{\boldmath$C_1$}. Here 
 \mbox{\boldmath$\Lambda$} is a flux limiter generalized to 2D as follows,
\begin{equation}
  \mbox{\boldmath$\Lambda$} =
\left(\begin{array}{cc}
  \Lambda_r & 0 \\
  0 & \Lambda_{\theta}\\
\end{array}\right),
\end{equation}
 whose components are given by $
 \Lambda_r = ({3\lambda^t})/({3 + \lambda^{t}|\nabla_r \psi_0|/ \psi_0})$, and 
$ \Lambda_{\theta} = ({3\lambda^t})
/({3 + \lambda^{t}|\nabla_{\theta} \psi_0|/ \psi_0})$,
where $\lambda^{t}$ is the transport mean free path (see Equation (A26) in \cite{bruenn85}), $\nabla_r$ and $\nabla_{\theta}$ represents $\partial / \partial
r$ and $\partial/ (r \partial \theta)$, respectively. With equation (\ref{psi1}), one can determine $\psi_0$ by solving the following zeroth angular moment equation of Boltzmann equation,  
\begin{equation}
\frac{1}{c}\frac{d}{d t} \psi_0+ 
\nabla\cdot \mbox{\boldmath$\psi_1$} + \frac{1}{c}
\nabla \cdot \mbox{\boldmath$v$}~\omega~\frac{\partial}{\partial \omega } \psi_0 =
j(1-\psi_0) - \frac{\psi_0}{\lambda^{a}} + A_0 \psi_0 + \mbox{\boldmath$B_0$}
\cdot \mbox{\boldmath$\psi_1$} + C_0,
\label{psi0}
\end{equation}
which is often referred to as a multi-group flux limited diffusion (MGFLD) equation for neutrinos (again see appendix A in \cite{bruenn85} for the definition of unmentioned variables). For the time evolution of transport, we solve the neutrino-matter coupling (equations 
(\ref{energy}) (the $p\nabla\cdot{\mbox{\boldmath{$v$}}}$ term is treated in an operator-splitting fashion), (\ref{ye_flow}), and (\ref{psi0}) implicitly with performing the Newton Raphson iteration until $\delta \psi_0$, $\delta Y_e$, and $\delta(e/\rho)$ converges to a certain value. More details for the numerical implementation of our code will be presented in the forthcoming paper. 

\section{Static background}
\begin{figure}[htb]
\begin{center}
\epsfxsize = 5.5 cm
\rotatebox{270}{\epsfbox{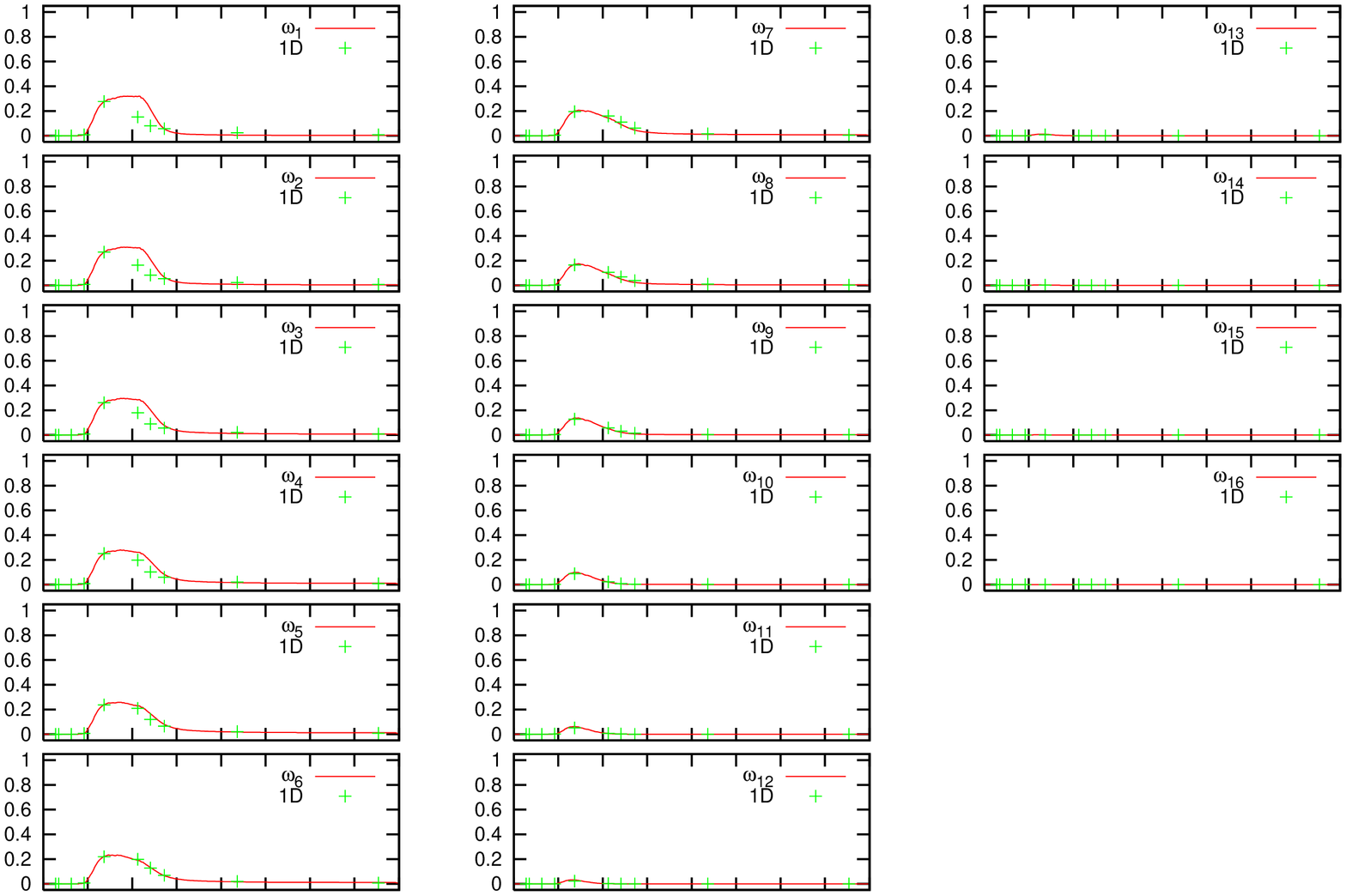}}
\epsfxsize = 5.5 cm
\rotatebox{270}{\epsfbox{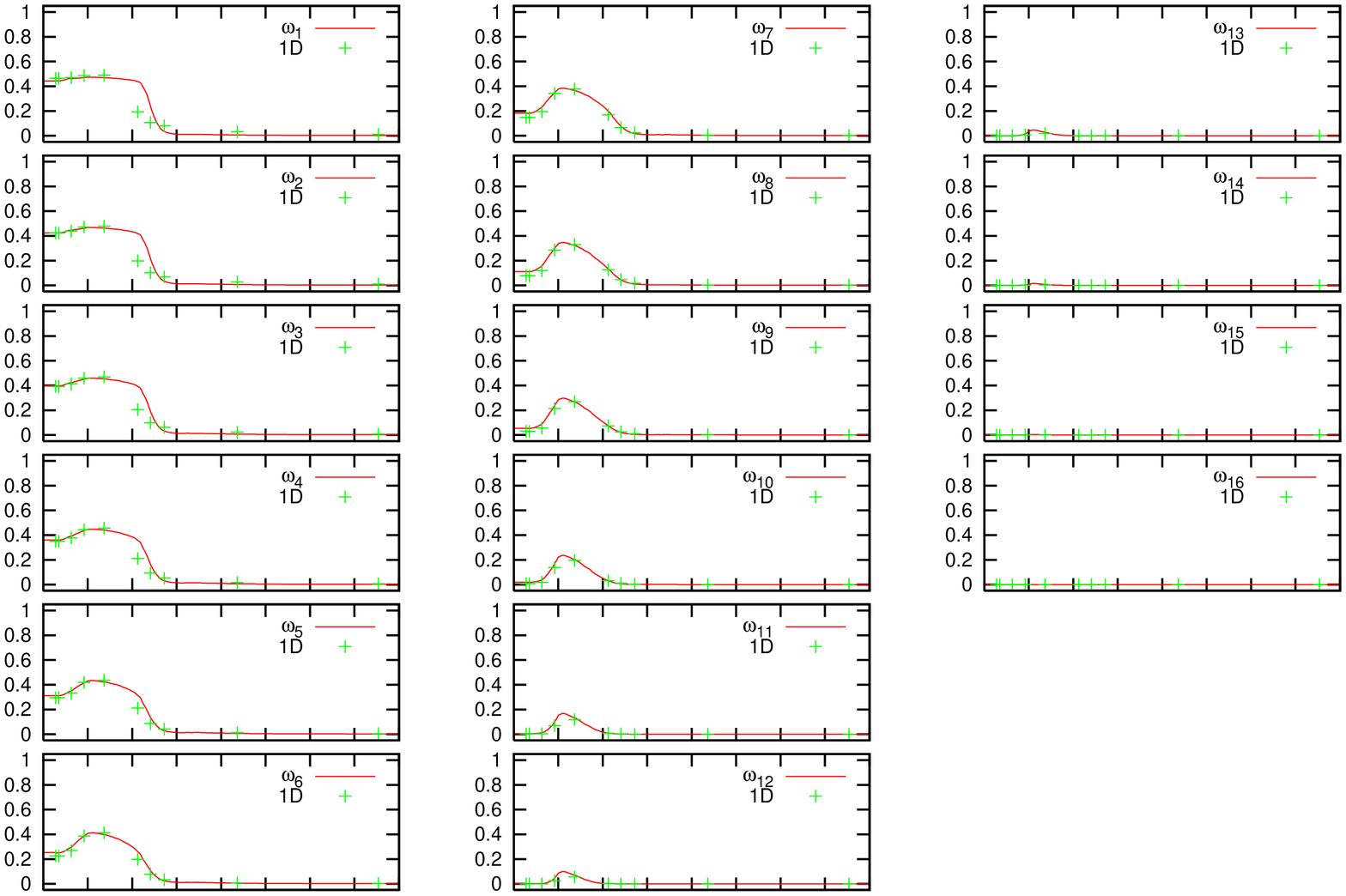}}
\end{center}
\caption{Anti-electron (left) and mu-tau (right) neutrino distribution functions at each energy bin as a function of the stellar radius (from 0 to 80 km, equidistantly plotted). Note in the figure that $\omega_{i} (i=1,16)$ represents the neutrino energy group and that ``1D'' represents the results by \cite{bruenn92}.}\label{nuebnumu}
\end{figure}

Before we apply the newly developed code to the magnetized and rotating supernova simulations, we need to verify the accuracy by the test calculations. Taking the profiles of the density, temperature,
and the electron fraction from Bruenn's 1D model after core bounce \cite{bruenn92} as a background,  we
calculate the neutrino distribution functions of each species ($\nu_e, 
\bar{\nu}_e, \nu_x, \bar{\nu}_x$) by our  2D computations and
compare them with the ones obtained in 1D simulation \cite{bruenn92}. We use the 16 energy mesh points which is logarithmically uniform and covers 0.95 - 255 MeV. 
In Figure \ref{nuebnumu}, the distribution functions of $\bar{\nu}_e$ and 
$\nu_{x}$ at each energy bin as a function of radius is shown.
One can see the good agreement with each other, except for the small 
discrepancies at $\sim$ 20 km for the lower energy bins. This is also the case
 for the electron neutrinos, which is not presented here. This may be mainly due to the difference of the numerical implementation of the neutrino transport, and of the employed equation of state. Since close comparison is made in the static background computation, we move on to apply the code to dynamical models.
\begin{figure}[hbt]
\begin{center}
\epsfxsize = 12.5 cm
\epsfbox{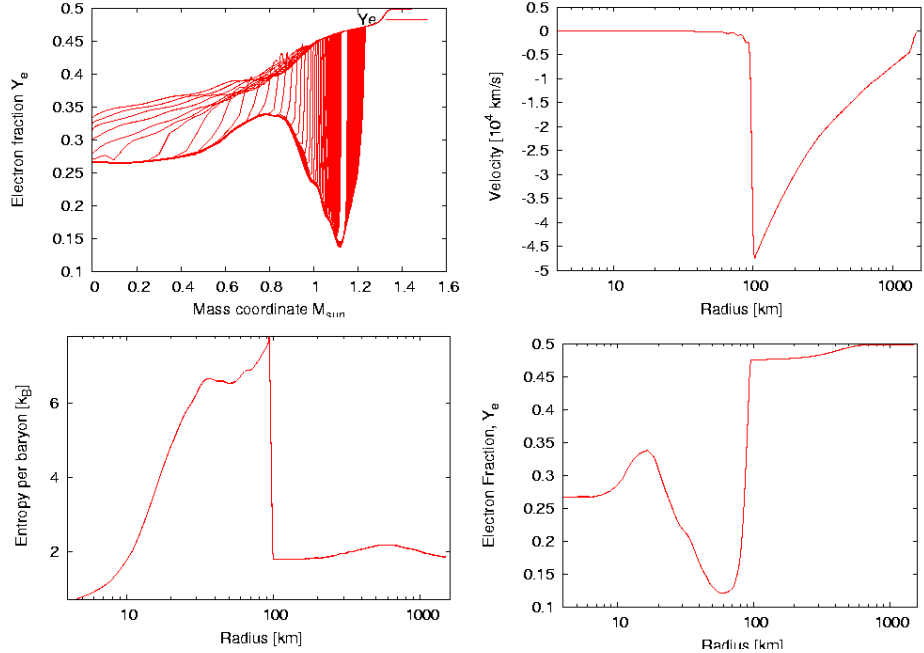}
\end{center}
\caption{Top left panel is the time evolution of $Y_e$ up to $\sim$40 msec after bounce (small blank is due to the data loss). Top right, bottom left, bottom right shows the velocity, entropy, $Y_e$ distribution at $10$ msec after bounce. Compare panels (a), (c), (d) in Figure 20 in \cite{Liebendoerfer_et_al_04}.}
\label{10msec}
\end{figure}
\section{Spherically symmetric simulation}
Here we examine the accuracy of our code by the spherically symmetric core-collapse simulations of 15 $M_{\odot}$ progenitor star (Woosley \& Weaver 1995). 
In Figure \ref{10msec}, the velocity (top right), 
entropy (bottom left), $Y_e$ (bottom right) distributions at 10 msec after bounce with the evolution of $Y_e$ up to $\sim 40$ msec after bounce (top left), are shown. 
The global profiles are well in good agreement with the previous spherically 
symmetric calculations \cite{1D}. Discrepancies with other calculations could be originated from the 
differences of employed microphysics, numerical treatment of neutrino 
transport, general relativistic effects, which seem rather common from models to models \cite{1D}. More details about the tests of our newly developed code will be presented soon in elsewhere,
 with its applications to the magnetorotational core-collapse simulations of massive stars. 
\section{Outlook}
Although $O(v/c)$ terms in the transport equations, which are one of the most important ingredients for determining the failure or success of the explosion \cite{buras}, are neglected in the current simulations, 
we are now implementing them, simultaneously running the MHD collapse models
 without to see the difference. It is better to employ more detailed neutrino 
microphysics. Steady progresses with the parallelization of the code are made to test the predictions in our earlier works.
\vspace{-0.5 cm}
\section*{Acknowledgments}
This work was supported in part by the Japan Society for
Promotion of Science (JSPS) Research Fellowships (K.K) and a
Grant-in-Aid for Scientific Research from the Ministry of
Education, Science and Culture of Japan through No.S 14102004,
No. 14079202, and No. 14740166.

\end{document}